\documentclass[a4paper,11pt]{article}
\usepackage[utf8]{inputenc}
\usepackage{jcappub}
\pdfoutput=1 
\usepackage[T1]{fontenc} 
\usepackage{verbatim}
\usepackage{color}
\usepackage{natbib}
\usepackage{soul}
\usepackage{physics}
\usepackage{placeins}
\usepackage{graphicx}
\usepackage{amsmath}
\usepackage{amsfonts}
\usepackage{subfigure}
\newcommand{\beq}{\begin{equation}}
\newcommand{\eeq}{\end{equation}}

\title{Is asymptotically safe inflation eternal?}

\date{December 2020}
\author[1,*]{J. Chojnacki,\note[*]{Corresponding author.}}
\author[1]{J. Krajecka,}
\author[1]{J. H. Kwapisz,}
\author[1,2,3]{O. Slowik,}
\author[1]{A. Strag}
\affiliation[1]{Faculty of Physics, University of Warsaw,
Pasteura 5, 02-093 Warsaw, Poland}
\affiliation[2]{Center for Theoretical Physics, Polish Academy of Sciences, Al. Lotników 32/46, 02-668 Warsaw, Poland}
\affiliation[3]{Faculty of Mathematics, Informatics and Mechanics, University of Warsaw, Banacha 2, 02-097 Warsaw, Poland}
\emailAdd{jr.chojnacki@student.uw.edu.pl}
\emailAdd{j.krajecka@student.uw.edu.pl}
\emailAdd{Jan.Kwapisz@fuw.edu.pl}
\emailAdd{oslowik@cft.edu.pl}
\emailAdd{a.strag@student.uw.edu.pl}
\abstract{Recently, based on swampland considerations in string theory, the (no) eternal inflation principle has been put forward. The natural question arises whether similar conditions hold in other approaches to quantum gravity. In this article, the asymptotic safety hypothesis is considered in the context of eternal inflation.\\
As exemplary inflationary models the SU(N) Yang-Mills in the Veneziano limit and various RG-improvements of the gravitational action are studied. The existence of UV fixed point generically flattens the potential and our findings suggest no tension between eternal inflation and asymptotic safety, both in the matter and gravitational sector in contradistinction to string theory. Moreover, the eternal inflation cannot take place in the range of applicability of effective field quantum gravity theory.\\
We employ the analytical relations for eternal inflation to some of the models with single minima, such as Starobinsky inflation, alpha-attractors, or the RG-improved models and verify them with the massive numerical simulations. The validity of these constraints is also discussed for a multi-minima model.}
\begin{document}
\maketitle
\flushbottom
\section{Introduction}
Our Universe consists of 4 fundamental forces. Three of these forces have been consistently described on the quantum level and combined into the Standard Model of particle physics. Only quantum gravity seems to be elusive and has not been fully described in terms of quantum theory. This is not only because gravity is power counting non-renormalizable but also due to the fact that direct quantum gravity regime cannot be accessed experimentally (for example an accelerator measuring the quantum gravity effects would have to be big as our Solar System). \\
In recent years an alternative strategy has been put forward, namely one formulates a fundamental quantum gravity theory and then tests, which of the low energy effective theories can be UV completed by this quantum gravity model. In string theory this goes under the name of swampland conjectures \cite{Vafa:2005ui,Ooguri:2006in}. Recently widely discussed is so-called de-Sitter conjecture \cite{Obied:2018sgi,Agrawal:2018own} which states that string theory cannot have de-Sitter vacua and is in tension with single field inflation \cite{Achucarro:2018vey,Kinney:2018nny}. There seems to be also a tension between standard S-matrix formulation of quantum gravity and existence of stable de-Sitter space \cite{Dvali:2013eja,Dvali:2014gua,Dvali:2017eba,Dvali:2020etd}. However it is not established whether asymptotic safety admits a standard S-matrix formulation \cite{Kwapisz:2020jux} due to fractal spacetime structure in the deep quantum regime \cite{Lauscher:2005qz,Lauscher:2005xz} In line of these swampland criteria the no eternal inflation principle has been put forward \cite{Rudelius:2019cfh}, see also the further discussions on the subject of eternal inflation \cite{Guth:2007ng,Johnson:2011aa,Lehners:2012wz,Leon:2017sru,Matsui:2018bsy,Wang:2019eym,Blanco-Pillado:2019tdf,Lin:2019fdk,Hohm:2019ccp,Hohm:2019jgu,Banks:2019oiz,Seo:2020ger}. \\
On the other hand, the theory of inflation is a well-established model providing an answer to problems in classical cosmology, such as the flatness problem, large-scale structures formation, homogeneity, and isotropy of the universe. A handful of models is in an agreement with the CMB observations. In the inflationary models, quantum fluctuations play a crucial role in primordial cosmology, providing a seed for the large-scale structure formation after inflation and giving a possibility for the eternally inflating multiverse. Initial fluctuations in the early universe may cause an exponential expansion in points scattered throughout the space. Such regions, rapidly grow and dominate the volume of the universe, creating ever-inflating, disconnected pockets. Since so far there is no way to verify the existence of the other pockets, we treat them as potential autonomous universes, being part of the multiverse.\\
In the light of this tension between string theory and inflationary paradigm \cite{Rudelius:2019cfh}, one can be interested in how robust are the swampland criteria for the various quantum gravity models. In accordance with the inflation theory, we anticipate that the dynamics of the universe are being determined by the quantum corrections to the general relativity stemming from the concrete UV model. The effective treatment led Starobinsky to create a simple inflationary model taking into account the anomaly contributions to the energy-momentum tensor. \\
As pointed out by Donoghue \cite{Donoghue:1994dn}
below the Planck scale, for quantum gravity one can safely take the effective field theory perspective. Yet these quantum gravity effects can be important below the Planck scale by the inclusion of higher dimensional operators. The gravitational constant $G_N$ has a vanishing anomalous dimension below the Planck scale and various logarithmic corrections to the $R^2$ have been considered, capturing the main quantum effects \cite{Codello:2014sua,Ben-Dayan:2014isa,Bamba:2014mua,Liu:2018hno}. Yet in order to get the correct 60 e-fold duration of inflationary period one has to push the scalar field value in the Einstein frame beyond the Planck mass \cite{Rudelius:2019cfh}. Furthermore most of these models do not possess a flat potential limit (either diverge or have a runaway solutions), suggesting that eternal inflation can be investigated only if one takes into account the full quantum corrections to the Starobinsky inflation. \\
In the effective field theory scheme, the predictive power of the theory is limited, as the description of gravity at transplanckian scales requires fixing infinitely many coupling constants from experiments. The idea of asymptotic safety \cite{Weinberg:1980gg} was introduced by Stephen Weinberg in 1978 as a UV completion of the quantum theory of gravity. The behavior of an asymptotically safe theory is characterized by scale invariance in the high-momentum regime. Scale invariance requires the existence of a non-trivial Renormalization Group fixed point for dimensionless couplings. There are many possible realizations of such non-trivial fixed point scenario, such as the canonical vs anomalous scaling (gravitational fixed point \cite{Reuter:1996cp,Souma:1999at,Lauscher:2001ya,Reuter:2001ag}) and one-loop vs two-loop contributions or gauge vs Yukawa contributions, see \cite{eichhorn2019asymptotically} for further details and \cite{Dupuis:2020fhh} for current status of asymptotically safe gravity.\\
The existence of an interacting fixed point and hence the flatness of the potential in the Einstein frame led Weinberg to discuss \cite{Weinberg_2010} cosmological inflation as a consequence of Asymptotically Safe Gravity, see also \cite{Reuter:2005kb,Reuter:2012id} for discussion of AS cosmology. Following this suggestion, we study two types of models.\\
The first type relies on the RG-improvement of the gravitational actions and is based on the asymptotic safety hypothesis that gravity admits a non-trivial UV fixed point. Since asymptotically safe gravity flattens the scalar field potentials \cite{Eichhorn:2017als}, one can expect that it will result in the eternal inflation for large enough initial field values. On the other hand, RG-improved actions can serve as a UV completion of the Starobinsky model. One should also note that asymptotically safe swampland has been vastly studied \cite{Shaposhnikov:2009pv,Zanusso:2009bs,Daum:2009dn,Folkerts:2011jz,Christiansen:2012rx,Wang:2015sxe,Eichhorn:2016esv,Grabowski:2018fjj,Kwapisz:2019wrl,Eichhorn:2017ylw,Eichhorn:2018whv,Eichhorn:2017muy,Eichhorn:2016vvy,Christiansen:2017cxa,Eichhorn:2018nda,Christiansen:2017gtg,Eichhorn:2019dhg,Eichhorn:2019yzm,Alkofer:2020vtb,Daas:2020dyo,Held:2020kze,Hamada:2020vnf,Reichert:2019car,Eichhorn:2020kca,Eichhorn:2020sbo,Hamada:2020mug,deBrito:2020dta}.\\
The other model relies on the non-trivial fixed point in the pure matter sector governed by the Yang-Mills dynamics in the Veneziano limit \cite{Litim_2014,Litim:2015iea}, see also \cite{Mann:2017wzh,Antipin:2018zdg,Molinaro:2018kjz,Wang:2018yer,Wang:2018yer}. In this model, we have uncovered a new type of eternal inflation scenario relying on tunneling to a false vacuum - in the opposite direction as it was considered in the old-inflation proposal \cite{Guth:1980zm}.\\
In contradistinction to string theory, the couplings in the asymptotic safety paradigm are predicted from the RG-flow of the theory and their fixed point values rather than as vacuum expectation values (vev's) of certain scalar fields. Hence, the asymptotically safe eternally inflating multiverses landscape is much less vast than the one stemming from the string theory, making these models much less schismatic \cite{Ijjas:2014nta}. Finally, let us note that asymptotic safety can argue for the homogeneous and isotropic initial conditions on its own using the finite action principle \cite{Lehners:2019ibe}.\\
Our work is organized as follows. In Chapter~\ref{sec:EI} we introduce the idea of eternal inflation and multiverse. We discuss necessary conditions for eternal inflation to occur based on the Fokker-Planck equation. In Chapter~\ref{sec:EM} we show, how the developed tools work in practice with the two popular inflationary models. Chapter~\ref{sec:EIinASM} is devoted to the presence of eternal inflation in Asymptotically Safe models. In Chapter~\ref{Sec:Discussion} the results are discussed and concluded.
\section{How inflation becomes eternal?}
\label{sec:EI}
In this section we discuss, under what circumstances the inflation becomes eternal. Our discussion follows closely \cite{Rudelius:2019cfh}.
\subsection{Fokker-Planck equation}
Consider a scalar field in the FLRW metric
\begin{align}
        S= \int d^4 \sqrt{-g} \left(\frac{1}{2}M_{Pl}^2 R+ \frac{1}{2}g^{\mu\nu}\partial_{\mu}\phi\partial_{\nu} \phi - V(\phi)\right),
\end{align}
with $\phi(t,\vec{x})=\phi(t)$ one obtains the following equations of motion:
\begin{align}
\ddot{\phi} + 3H\phi + \frac{\partial V}{\partial \phi} = 0, \quad H^2 M_P^2 = \frac{1}{3} \left(\frac{1}{2} \dot{\phi}^2 + V(\phi)\right),
\end{align}
which in the slow-roll approximation become \cite{Kwapisz:2019cxq}:
    \begin{align}
    \label{eq:slowroll}
    3H\dot{\phi}+\frac{\partial V}{\partial\phi} \approx0, \quad & H^2M_{Pl}^2 \approx \frac{1}{3}V\left(\phi\right ).
    \end{align}
 Inflation ends once one of the so-called slow-roll parameters becomes of order one
 \begin{align}
\label{Inflconditions}
\epsilon \simeq \frac{M_P^2}{2} \left(\frac{V_{,\phi}}{V}\right)^2, \quad
\eta   \simeq M_P^2 \frac{V_{,\phi\phi}}{V},
\end{align}
 and enters the oscillatory, reheating phase. 
The standard treatment of eternal inflation relies on the stochastic inflation approach \cite{Linde:1991sk}. One splits the field into classical background and short-wavelength quantum field
\begin{align}
    \phi\left(t,\Vec{x}\right )=\phi_{cl}\left(t,\Vec{x}\right )+\delta\phi\left(t,\Vec{x}\right ).
\end{align} 
Due to the fact that action is quadratic in the fluctuations, their spatial average over the Hubble volume is normally distributed. Hence from now on we shall assume that both background and fluctuations are homogeneous, which is the standard treatment of eternal inflation (if not otherwise specified). In the large e-fold limit, equation of motion for the full field takes form of slow-roll equation with additional classical noise term \cite{Rudelius:2019cfh,Kiefer:1998qe,Kiefer:2008ku}, known as the Langevin Equation:
\begin{align}
\label{Langevin}
    3H\dot{\phi}+\frac{\partial V}{\partial\phi}  =N\left(t\right ), 
\end{align}
where $N\left(t\right )$ is a Gaussian distribution with mean equal 0 and variance $\sigma=\frac{H^3t}{4\pi^2}$ \cite{Linde_1992}.
Then the probability density of the inflaton field is then given by the Fokker-Planck equation \cite{Rudelius:2019cfh}:
\begin{align}
\label{Planck-Fokker}
\dot{P}[\phi,t]=\frac{1}{2}\left(\frac{H^3}{4\pi^2}\right)
\frac{\partial^2 P[\phi,t]}{\partial \phi \partial \phi}+\frac{1}{3H}\partial_i\left(\partial^i V\left(\phi\right) P[\phi,t]\right),
\end{align}
where 
$\dot{P}[\phi,t]:=\frac{\partial}{\partial t} P[\phi,t]$.
\subsection{Analytic solutions}
To understand better the Fokker-Planck equation, let us now briefly discuss the analytical solutions.\\
\paragraph{Case 1. Constant potential}
 \begin{align}
V\left(\phi\right ) = V_0,
 \end{align}
the Fokker-Planck equation reduces to
\begin{align}
    \dot{P}[\phi,t]=\frac{1}{2}\left(\frac{H^3}{4\pi^2}\right)
\frac{\partial^2 P[\phi,t]}{\partial \phi \partial \phi},
\end{align}
furthermore $H^2 =\textrm{const}$ by the Friedman equations.
 Then the Fokker Planck equation reduces to the standard heat equation, which
has a solution given by a Gaussian distribution:
\begin{align}
    \label{gauss}
    P[\phi,t] = \frac{1}{\sigma\left(t\right )\sqrt{2\pi}}\exp\left[-\frac{\left(\phi - \mu\left(t\right )\right )^2}{2\sigma\left(t\right )^2}\right],
\end{align}
 with
\begin{align}
    \mu\left(t\right ) = 0, \quad & \sigma^2\left(t\right ) = \frac{H^3}{4\pi^2}t.
\end{align}
A delta-function distribution initially centered at $\phi$ = 0 will remain centered at $\phi = 0$ for all time. It will however, spread out by the amount $\sigma \left ( t = H^{-1}\right ) = H/2\pi$ after a Hubble time. This represents the standard “Hubble-sized” quantum fluctuations that are well-known in the context of inflation, famously imprinted in the CMB and ultimately seeding the observed large-scale structure. \\
\paragraph{Case 2. Linear potential}
For the linear hilltop model the potential is given by
\begin{align}
    \label{eq:linearhilltop}
    V\left(\phi\right ) = V_0 - \alpha \phi.
\end{align}
Fokker-Planck equation is analogously solved by the Gaussian distribution (\ref{gauss}) with:
\begin{align}
\label{linearsolution}
    \mu\left(t\right ) = \frac{\alpha}{3H}t, \quad & \sigma^2\left(t\right ) = \frac{H^3}{4\pi^2}t.
\end{align}
The time-dependence of $\mu\left(t\right )$ is due to  the classical rolling of the field in the linear potential.
The time-dependence of $\sigma^2\left(t\right )$ is purely due to Hubble-sized quantum fluctuations, and it precisely matches the result in the constant case. In general, for a linear and quadratic potential the equation simplifies to  the heat equation, hence the solutions are Gaussian. Furthermore, if the potential is asymptotically flat, a finite limit at infinity exists. One may employ a series expansion around this point at infinity, approximating the potential up to the linear term. It is then expected that the probability density is approximately Gaussian (\ref{linearsolution}). In the next section, we describe in detail how the Gaussian distribution causes the inflaton to decay exponentially.  
\subsection{Eternal inflation conditions}
Given an arbitrary field value $\phi_c$, one can ask what is the probability that quantum field $\phi=\phi(t)$ is above this value:
\begin{align}
\label{phi_c}
\mathrm{Pr}[\phi>\phi_c,t]=\int^{\infty}_{\phi_c}d{\phi}P[\phi,t].
\end{align}
Since the distribution is Gaussian, then for $\phi_c$ large enough the $Pr[\phi>\phi_c,t]$ can be approximated by an exponential decay:
\begin{align}
\label{eq:gamma}
   \mathrm{Pr}[\phi > \phi_c,t] \approx C(t)\exp(-\Gamma t),
\end{align}
where $C(t)$ is polynomial in $t$ and all of the dependence on $\phi_c$ is contained in $C(t)$. Then it seems that inflation cannot last forever since
\begin{align}
     \lim_{t\to \infty}\mathrm{Pr}[\phi>\phi_c,t] =0.
\end{align}
However, there is an additional effect to be included: expansion of the universe during inflation. The size of the universe depends on time according to:
\begin{align}
     U\left(t\right )=U_0 e^{3Ht},
\end{align}
where $U_0$ is the initial volume of the pre-inflationary universe.
One can interpret the probability $ Pr[\phi>\phi_c,t]$ as fraction of the volume $U_{inf}\left(t\right )$ still inflating, that is: 
\begin{align}
\label{eq:eternal}
U_{inf}\left(t\right )=U_0e^{3Ht}  Pr[\phi>\phi_c,t],
\end{align}
then in order for the Universe to inflate eternally, the positive exponential factor $3H$ in Eq.~(\ref{eq:eternal}) and the negative exponential factor $-\Gamma$ in (\ref{eq:gamma}) must satisfy: 
\begin{align}
    3H>\Gamma.
\end{align}
We shall illustrate this general property on an example of linear potential. Evaluating the integral for probability density, in the linear case gives:
\begin{align}
Pr[\phi>\phi_c,t]=\frac{1}{2} \textrm{erfc}\left({\frac{\frac{\alpha}{3H}t-\phi_c}{\frac{H}{2\pi}\sqrt{2Ht}}}\right ).
\end{align}
The error function may be approximated by an exponential:
\begin{align}
Pr[\phi>\phi_c,t]=C\left(t\right )\textrm{exp}\left( -\frac{4\pi^2\alpha^2}{18H^5}t \right ),
\end{align}
where $C\left(t\right )$ is power-law in $t$
and $\phi_c$ vanished from the final approximation of the probability, which is a generic feature. By comparing the exponents we can check, whether $U_{inf}$ will grow or tend to zero. The condition for eternal inflation to occur becomes:
\begin{align}
     3H>\frac{4\pi^2\alpha^2}{18H^5}.
\end{align}
For linear potential $\alpha=V'\left(\phi\right )$ using the slow-roll equations equation (\ref{eq:slowroll}), above condition can be rewritten:
\begin{align}
\label{eq:EternalCondition1}
     \frac{|V'|}{V^{\frac{3}{2}}}<\frac{\sqrt{2}}{2\pi} \frac{1}{M^2_{Pl}}.
\end{align}
This can be interpreted as quantum fluctuations dominating over classical field rolling. For linear potential, this is satisfied for a large $\phi$. Similarly, the second condition for the eternal inflation may be derived from the quadratic hilltop potential:
\beq
\label{eq:EternalCondition2}
-\frac{V''}{V}<\frac{3}{M^2_{Pl}}.
\eeq
Further necessary conditions on p-th derivative with $p>2$ have been derived in \cite{Rudelius:2019cfh} and give:
\begin{align}
    [-\textrm{sgn}\left(\partial^p  V\right)]^{p+1}\frac{|\partial^p V|}{V^{(4-p}/2}< \mathcal{N}_p M_{Pl}^{p-4},
\end{align}
where $\mathcal{N}_p\gg 1$ is numerically determined coefficient. Eternal inflation can be understood as a random walk of a field and a diffusion process on top of the classical motion \cite{Vilenkin,Guth:2000ka,Guth:2007ng}.
In order to cross check the formulas (\ref{eq:EternalCondition1}, \ref{eq:EternalCondition2}) the numerical simulation has been developed. To reconstruct the probability distribution one simulates the discretized version of equation (\ref{Langevin}):
\begin{align}
\label{DiscretLangevin}
 \phi_n=\phi_{n-1}-\frac{1}{3H}V'\left(\phi_{n-1}\right )\delta t +\delta \phi_q \left(\delta t\right ),    
\end{align}
with $\delta \phi_q \left(\delta t\right )$ being random number taken from the gaussian distribution with mean equal zero, and variance $\frac{H^3}{4\pi^2}\delta t$. We further assume the Hubble parameter to be constant and respecting the slow-roll regime $H=\frac{1}{M_{Pl}}\sqrt{\frac{V(\phi_0)}{3}}$, where $V(\phi_0)$ is a value of the potential at the start of the simulation. We verified, that the change of $H$ caused by the field fluctuation does not affect the conclusions for eternal inflation. The simulation starts at the user-given value $\phi_0$ and follows the Langevin discretized equation (\ref{DiscretLangevin}).\\ If the inflation occurs, the corresponding timestep $t_n$ is added to a list. This happens while the slow-roll conditions are satisfied, meaning $\epsilon(t_n)$ and $\eta (t_n)$ are smaller than one. Violation of one of these conditions resets the simulation. However, the list containing information about time $t_n$ of the ongoing inflation is stored in the memory. This large time list is appended in the similar way each evolution. Its size may be estimated by $N\frac{T_{c}}{\delta t}$, where $N$ is the total number of simulations and $T_{c}$ is the time of the classical slow-roll inflation starting at $\phi_0$.\\ It is important to stress, that the duration of a particular evolution may be too long to compute in any practical time. We employ a large timeout ending the evolution. This is a good approximation for our purposes. Finally, the list containing information about every timestep at which inflation was ongoing in $N$ independent simulations is sorted in an ascending order. A normalized histogram with 1000 equal-width bins is created from the list. The number of counts is related to the probability of ongoing inflation, while the bins correspond to inflationary time. The field's evolution supports the Fokker-Planck result (\ref{eq:gamma}). In the slow-roll regime, the inflaton decays exponentially with decay parameter $\Gamma$. This is true for every numerically investigated potential in this work. In order to recognize the eternally inflating models we search for such initial value of the field $\phi_0$ that $\Gamma<3H$. In the numerical analysis we eliminate $\phi_c$ in equation (\ref{phi_c}) and instead check for the slow-roll conditions violation at each step.
\subsection{Tunneling and eternal inflation}
\label{SecTunneling}
\begin{figure}[t!]
\hfill
\subfigure[]{
\label{fig:Vacua1}
\includegraphics[width=0.45\textwidth ]{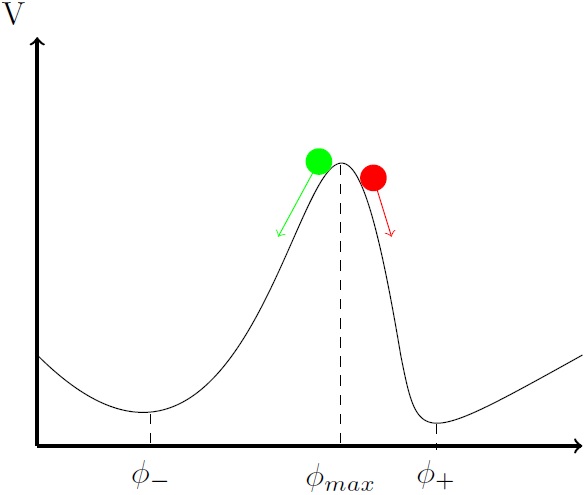}}
\hfill
\subfigure[]{
\label{fig:Vacua2}
\includegraphics[width= 0.45\textwidth]{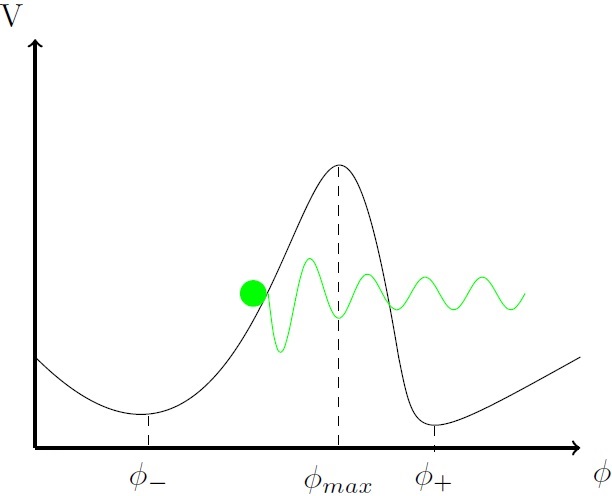}}
\hfill
\caption{Left: field initially placed at the maximum of the potential may decay towards one of the two vacua: at $\phi_{-}$ with probability $p_{-}$ and at $\phi_{+}$ with probability $p_{+}$.\newline
Right: field initially placed at $\phi_0<\phi_{max}$ may tunnel trough the barrier towards $\phi_{+}$ with probability $p_{+}(\phi_0)$. Analogous tunneling from "plus" to "minus" side is also possible.}
\label{Fig:1}
\end{figure}

Most of the inflationary potentials are of the single-minimum type such as Starobinsky inflation and alpha-attractors. There are however, potentials which are of type depicted on figure \ref{Fig:1} and possess a various minima. In such models, the inflation can become eternal due to tunnelling to the false vacua. When the vacua are degenerate enough, the tunneling dominates over quantum uphill rolling. The tunnelling goes in the opposing direction to the old inflation scenario \cite{Guth:1980zm} as showed on figure \ref{fig:Vacua2}. As it will turn out this is the dominant effect for the model discussed in section \ref{Sannino}.
The eternal inflation mechanism discussed in the previous sections relies on the local shape of the potential and cannot provide an accurate description in that case. In order to quantitatively derive predictions for this new effect, we shall rely on the first passage formalism 
 \cite{Vennin_2015,Noorbala_2018} instead, and apply it to the eternal inflation considerations. 
 Given the initial value of the field $\phi_0$ being between $\phi_{-}$ and $\phi_{+}$, the probability that it reaches $\phi_{+}$ before $\phi_{-}$ and $\phi_{-}$ before $\phi_{+}$, denoted respectively $p_{+}(\phi_0)$ and $p_{-}(\phi_0)$, obeys the following equation:
\begin{align}
vp''_{\pm}(\phi)-\frac{v'}{v}p'_{\pm}(\phi)=0,    
\end{align}
with initial conditions: $p_{\pm}(\phi_\pm)=1$, $p_\pm(\phi_\mp)=0$, where $v=v(\phi)$ is the dimensionless potential:
\begin{align}
    v(\phi):=\frac{V(\phi)}{24 \pi^2 M_{Pl}^4}.
\end{align}
The analytical solution is:
\begin{align}
\label{AnalP}
    p_\pm(\phi_0)=\pm \frac{\int^{\phi_0}_{\phi_\pm} e^{-\frac{1}{v(\phi)}} d\phi}{\int^{\phi_{+}}_{\phi_{-}}e^{-\frac{1}{v(\phi)}}d\phi}.
\end{align}
One may also define the probability ratio $R$:
\begin{align}
\label{ProbabilityR}
    R(\phi_0):=\frac{p_{+}(\phi_0)}{p_{-}(\phi_0)} =\frac{\int^{0}_{\phi_{-}} e^{-\frac{1}{v(\phi)}}d\phi}{\int^{\phi_{+}}_{\phi_0}e^{-\frac{1}{v(\phi)}}d\phi}.
\end{align}
The above integrals may be evaluated numerically. However, if the amplitude of $v(\phi)$ is much smaller than 1, the term $e^{-1/v(\phi)}$ will be extremely small, possibly exceeding machine precision in the computation. \\
Yet, one can use the steepest descent approximation. Consider a potential with the field located initially on the local maximum $\phi_{max}$ with a minimum on each side, as depicted on figure \ref{fig:Vacua1}. Then the probability ratio $R$ may be evaluated approximately, where the leading contributions to (\ref{ProbabilityR}) come from the values of the field in the neighborhood of  $\phi_{max}$.
$p_{+}$ and $p_{-}$ gives a probability of evolution realised respectively by the red and the green ball. 
We get:\footnote{For the details of the calculation consult \cite{Noorbala_2018}.}
\begin{align}
\label{Rmaximum}
    R(\phi_{max})\approx 1-\frac{2}{3}\frac{\sqrt{2}}{\pi} \frac{v(\phi_{max}) v'''(\phi_{max})}{|v''(\phi_{max})|^{3/2}}.
\end{align}
In this regime, probability of descending into each of the minima $\phi_{-}$ and $\phi_{+}$ is similar, giving $|1-R|\ll 1$. It is possible to start the inflation in the subset of $[\phi_{-},\phi_{+}]$ that would lead to the violation of the slow-roll conditions and tunnel through the potential barrier to the sector dominated by eternal inflation, schematically shown on figure \ref{fig:Vacua2}. We further analyze this possibility in Sec.~\ref{tunneling section} for a particular effective potential with two vacua, stemming from an asymptotically safe theory. We use equation ~(\ref{Rmaximum}) to find the dependence of $R$ on the parameters of the theory and verify the result with direct numerical simulation of the Langevin Equation (\ref{Langevin}) given set of parameters.
\section{Exemplary models}
\label{sec:EM}
In this section we show the basic application of the conditions (\ref{eq:EternalCondition1}, \ref{eq:EternalCondition2}) to simple effective potentials stemming from the $\alpha$-attractor models and the Starobinsky inflation.
\subsection{Alpha-attractor models}
\label{AlfaAttractorSection}
\begin{figure}[t!]
\hfill
\subfigure[]{
\label{fig:AlfaAttractorPotential}
\includegraphics[width=0.48\textwidth]{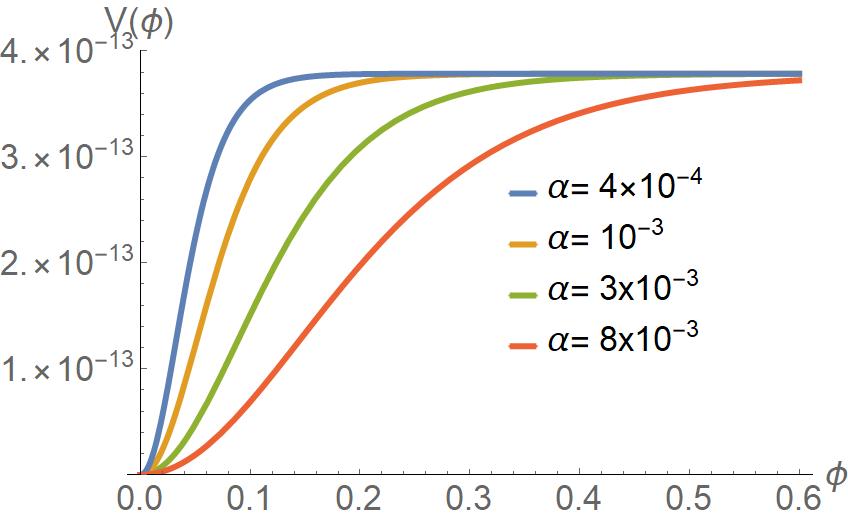}}
\hfill
\subfigure[]{
\label{fig:AlfaAttractorEternalInflation}
\includegraphics[width=0.48\textwidth]{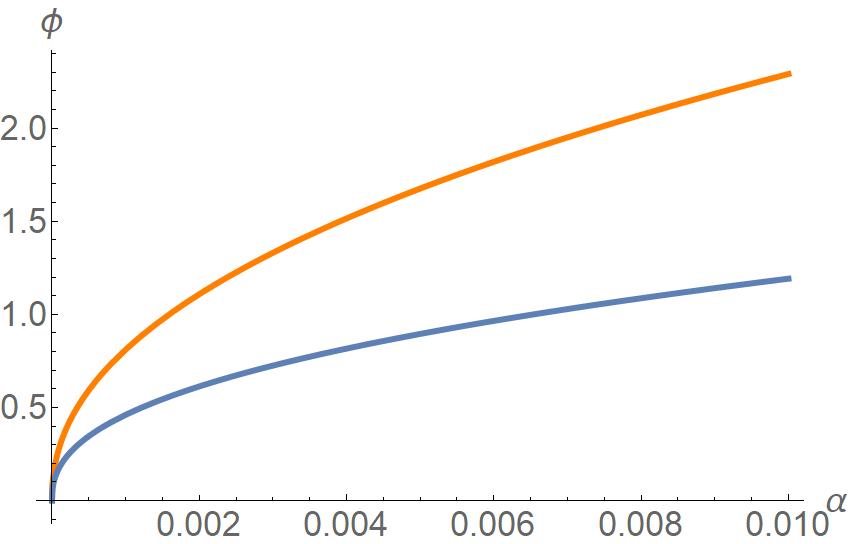}}
\hfill
\caption{Left: the T-model potential with $n=1$ and various $\alpha$ were depicted.\newline
Right: plot of the initial value $\phi_0$ necessary for 60 e-folds, as a function of $\alpha$ (blue), as well as the lowest initial value $\phi_{EI}$ of the field, at which the eternal inflation "kicks in" (yellow).}
\end{figure}

We start our investigation with the $\alpha$-attractor models \cite{Carrasco_2015}, a general class of the inflationary models, originally introduced in the context of supergravity. They are consistent with the CMB data, and their preheating phase has been studied on a lattice in \cite{Krajewski_2019}. The phenomenological features of these models are described by the Lagrangian:
\begin{align}
    \frac{1}{\sqrt{-g}}\mathcal{L}_T=\frac{1}{2} R- \frac{1}{2}\frac{\partial \phi^2}{\left( 1-\frac{\phi^2}{6\alpha}\right)^2}-V(\phi).
\end{align}
Here, $\phi$ is an inflaton and $\alpha$ can take any real, positive value.
At the limit $\alpha\xrightarrow{}\infty$ the scalar field becomes canonically normalized, and the theory coincides with the chaotic inflation. Canonical and non-canonical fields are related by the transformation:
\begin{align}
    \phi=\sqrt{6\alpha}\tanh{\frac{\varphi}{\sqrt{6\alpha}}}.
\end{align}
We further consider T-models, in which the potential of canonically normalised field is given by:
\begin{align}
\label{AlfaV}
    V(\phi)=\alpha \mu^2 \tanh^{2n}{\frac{\varphi}{\sqrt{6\alpha}}},
\end{align}
where parameter $\mu$ is of order $10^{-5}$. The shape of the potential for $n=1$ was plotted on figure \ref{fig:AlfaAttractorPotential}.
At the large $\phi$ the potential (\ref{AlfaV}) is asymptotically flat, this creates the possibility for eternal inflation to occur. Using the first condition (\ref{eq:EternalCondition1}) we have verified, that generally the space is eternally inflating for all initial values of the $\phi$, above certain $\phi_{EI}$. The second eternal condition (\ref{eq:EternalCondition2}) as well as the higher order conditions are satisfied for almost all values of $\phi_0$ above $0$, providing no new information. This is a generic feature for all of the models we investigate. \\For every $\alpha$, $\phi_0$ necessary to produce 60 e-folds is safely below $\phi_{EI}$. We have found $\phi_0$ by solving slow-roll equation numerically. It is shown on figure \ref{fig:AlfaAttractorEternalInflation}.
The time of inflation larger than 60 Planck-times is unlikely according to the Planck Collaboration data \cite{refId0}. The values of $\phi_{EI}$ change only slightly with $n$. We may therefore conclude, that $\alpha$-attractor models are consistent with the beginning of "our" pocket universe. However, it is not inconceivable that the field fluctuations in other parts of the early universe had values $\phi_0>\phi_{EI}$, driving the eternal inflation.
\FloatBarrier
\subsection{Starobinsky model}
\begin{figure}[t!]
\label{starysample}
\hfill
\subfigure{\includegraphics[width=0.49\textwidth]{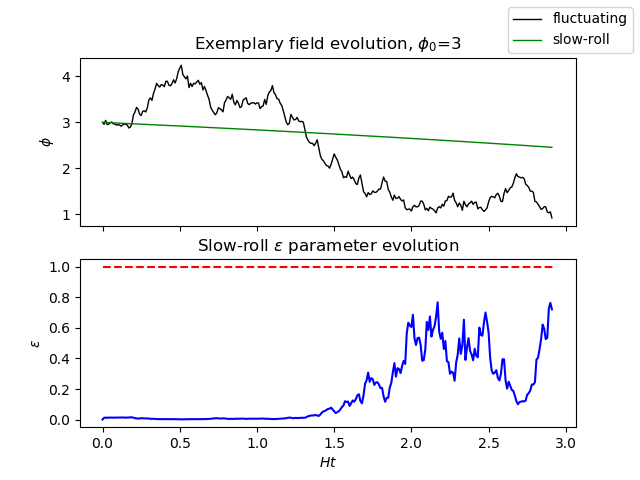}}
\hfill
\subfigure{\includegraphics[width=0.49\textwidth]{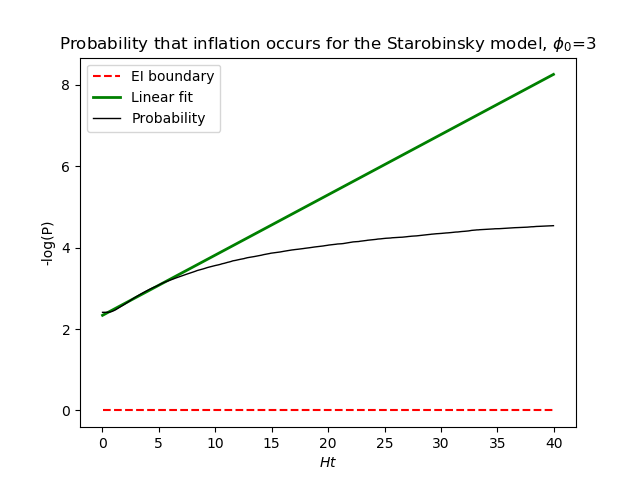}}
\hfill
\caption{Left: exemplary field evolution for the Starobinsky model has been plotted. Green plot shows the solution to the classical slow-roll equation, and the black plot is the Langevin solution. The inflation ends, when the slow-roll parameter reaches 1. Values of the field are given in $M_{Pl}$.\newline
Right: the time dependence of the probability, that inflation still occurs. In the slow-roll regime, the probability decays exponentially. Linear fit slope, the decay rate is around $\Gamma=0.15$. Red, dashed line denotes eternal inflation threshold with slope $3H$ of order $10^{-5}$. Since $3H<\Gamma$, initial condition $\phi_0=3$ $M_{Pl}$ is an example of non-eternally inflating universe.}
\end{figure}

Solutions stemming from Einstein-Hilbert action predict initial singularity. In 1980 Starobinsky proposed a model \cite{1980PhLB...91...99S}, where pure modified gravitational action can cause non-singular evolution of the universe, namely:
\begin{align}
\label{StaryLagrangian}
    S = \frac{1}{2}\int \sqrt{|g|}d^4x\left(M^2_pR + \frac{1}{6M^2}R^2\right ),
\end{align}
this can be rewritten to the effective potential form, with:
\begin{align}
\label{StaryPotential}
    V\left(\phi\right )=V_0\left(1-\exp\left( -\sqrt{\frac{2}{3}}\frac{\phi}{M_{Pl}}\right )\right )^2.
\end{align} 
The inflation begins on a plateau at large $\phi$. The field rolls towards a minimum at $\phi=0$, where the oscillatory reheating phase occurs. It has been estimated from the CMB data, that during the inflation the volume of the universe has grown by approximately 60 e-folds. 
This corresponds to the initial condition $\phi_0=5.5$ $M_{Pl}$, without taking into account quantum gravity effects. It is possible to perturbatively recover information about the shape of the potential from the CMB, for details see \cite{Lidsey_1997}. Amplitude of the scalar power spectrum $A_s=2\times10^{-9}$ fixes the value $V_0=8.12221\times10^{-11}$ $M^4_{Pl}$, via the relation:
\begin{align}
\label{cmb}
V_0= 24 \pi^2 \epsilon(\phi_0)A_s.    
\end{align}
Applying the analytical eternal inflation conditions (\ref{eq:EternalCondition1}, \ref{eq:EternalCondition2}) to the Starobinsky potential, the initial value of the field, above which the eternal inflation occurs, has been estimated to be $\phi_0=16.7$ $M_{Pl}$. It has been found, that decay rate $\Gamma$ decreases approximately exponentially with $\phi_0$. Our numerical simulation confirms the analytical prediction discussed in \cite{Rudelius:2019cfh} within a sample of 10000 simulations. They are performed as described in the previous section. An exemplary numerical evolution of the Langevin equation (\ref{Langevin}) is shown in the figure \ref{starysample}. The linear fit in the early times shows the exponential decay of the inflation in the slow-roll regime.\\ Nevertheless, for eternal inflation scenario and realistic phenomenology, this model requires
the transplanckian values of the fields in order to reproduce the correct tensor to scalar ratio $r$, amplitudes and spectral tilt $n_s$. Hence the Starobinsky inflation will be affected, possibly invalidated by the quantum gravity fluctuations. The leading log corrections have been studied in \cite{Liu:2018hno}, which we as well study in the context of eternal inflation. Yet, due to the large field values required for eternal inflation to occur, one should seek a theory predictive up to an arbitrary large energy scale, which we discuss in the next section.
\section{Eternal inflation in asymptotically safe models}
\label{sec:EIinASM} 
In this chapter, as a warm up we study effective corrections to the Starobinsky inflation, providing a different behavior at the large field values.
Later we show, that RG-improvement of Starobinsky model proposed in \cite{Bonanno:2015fga}, see also \cite{Bonanno:2017pkg, Platania:2020lqb} for review, closely related to the $R+R^2$ renormalizable Fradkin-Tseytlin gravity \cite{Fradkin:1981hx} produces a branch of inflationary potential entirely dominated by eternal inflation. We find the initial values of the inflaton such that inflation becomes eternal, for the remaining branch, as a function of theory parameters. Finally we show, that the possibility of tunneling through the potential barrier present in \cite{Nielsen:2015una} becomes a new mechanism for eternal inflation. In all of the asymptotically safe inflationary theories, eternal inflation is present as a consequence of asymptotic flatness of the effective potential.
\subsection{Quantum corrections to the Starobinsky model}
\begin{figure}[t!]
\hfill
\subfigure[]{
\label{fig:RStarobinskya}
\includegraphics[width=0.48\textwidth]{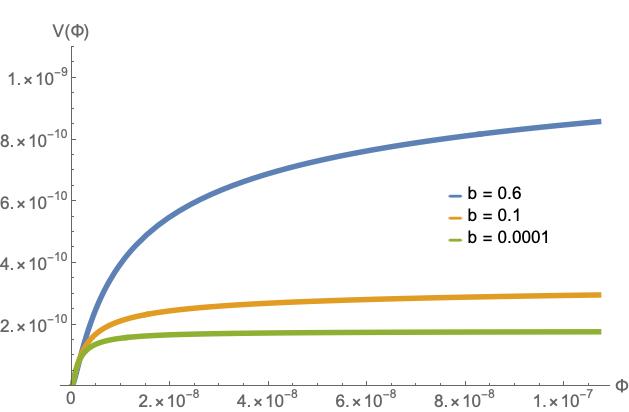}}
\hfill
\subfigure[]{
\label{fig:RStarobinskyb}
\includegraphics[width=0.48\textwidth]{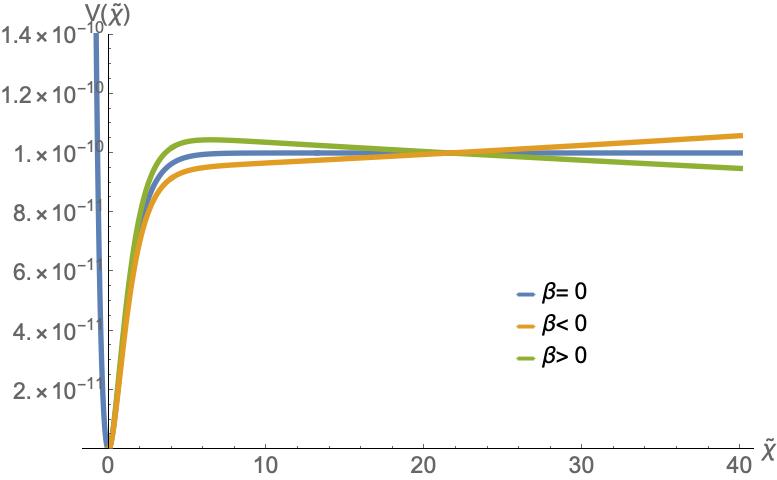}}
\hfill
\caption{Left: Plots of the potential (\ref{Veff_RS}) for various $b$ parameters.\newline
Right: Potential (\ref{smallb}) in dependence of $\beta$ parameter values. Approach towards infinity in case of $\beta > 0$ is visible.}
\end{figure}

Below Planck scale the gravitational constant $G_N$ has a vanishing anomalous dimension and the $R^2$ has a coefficient that runs logarithmically \cite{Demmel:2015oqa} (this comes from the fact that $R^2$ is dimensionless in $4$ dimensions). Hence, one can motivate various quantum corrected inflationary models, such as \cite{Codello:2014sua,Ben-Dayan:2014isa,Bamba:2014mua,Liu:2018hno}. In particular, the leading-log corrections to the Starobinsky model are given by \cite{Liu:2018hno}:
\begin{align}
    \label{L_RS}
    \mathcal{L}_{eff} = \frac{M_{Pl}^{2} R}{2} + \frac{\frac{a}{2}R^2}{1+b \ln(\frac{R}{\mu^2})}+\mathcal{O}(R^3).
\end{align}
In order to find the Einstein frame potential for this model a few steps need to be taken. First, we use the conformal transformation \cite{Kaiser:2010ps}.
Then, following the next transformation for the Ricci scalar and the metric determinant we get the Einstein frame action:
\begin{align}
    \label{EffActionFinal_RS}
       S = \int d^4 x \sqrt{-g_E} \left[\frac{M_{Pl} ^2}{2} R_E - \frac{1}{2} g_E ^{\mu \nu} \left(\partial_\mu \phi_E\right) \left(\partial_\nu \phi_E\right) - V_E (\phi_E) \right],
\end{align}
which depends on the sought potential $V_E$, that can be further obtained:\footnote{Here, the effective action differs from \cite{Liu:2018hno} due to the introduction of auxiliary numerical parameter\\ $e\approx 2.81$. Nevertheless, the dynamics stemming from each of the potentials are equivalent.}
\begin{align}
    \label{Veff_RS}
    V_E (\Phi) =\frac{M_{Pl}^4}{2} \frac{a \Phi^2 \left(1+b \ln \left(\frac{\Phi}{\mu^2}\right)\right)^2 \left(1+b \ln \left(\frac{\Phi}{e\mu}^2\right)\right)}{\left\{M_{p}^{2} \left(1+b \ln \left(\frac{\Phi}{\mu^2}\right)\right)^2 + 2 a \Phi \left(1+b \ln \left(\frac{\Phi}{\sqrt{e}\mu^2}\right)\right)\right\}^2},
\end{align}
with $\phi_E$ given by
\beq
F(\phi_E)=M_{Pl}^2\exp\left(\sqrt{\frac{2}{3}}\frac{\phi_E}{M_{Pl}}\right)=M_P^2 + \frac{a\Phi[2-b+2b\ln (\Phi/\mu^2)]}{[1+2b\ln (\Phi/\mu^2)]^2},
\eeq
yet the transformation between $\Phi$ and $\phi_E$ is non-invertible. By taking into account the COBE normalization, we can treat $b$ as a free parameter and fix $a(b)$. For $b=0$ one obtains the usual Starobinsky model, and for $b \ll 1 $ one gets the model discussed in \cite{Ben-Dayan:2014isa} $R^2(1+\beta \ln R)$
with the potential given by Lambert W function (so the same as for model discussed in section \ref{Sannino}) and approximated in the limit $\beta \ll 1$ as
\beq
\label{smallb}
V \approx \frac{V_s}{1+b/(2\alpha)+ \beta/\alpha\ln[(e^{\tilde{\chi}}-1)/2\alpha]},
\eeq
where $V_s$ is the Starobinsky potential, $\tilde{\chi}$ is the Einstein frame field and $\alpha(\beta)$, where we have kept the original notation. \\
From the plots~\ref{fig:RStarobinskya}, \ref{fig:RStarobinskyb}, one can see that both of the models should give similar inflationary observables as in the Starobinsky inflation. On the other hand the eternal inflation in this models will be quite different. These models for $\beta<0$ and $b>0$ have the potentials that are non-flat for large field values, while for $\beta>0$ potential depicted on Figure \ref{fig:RStarobinskyb} have the runaway behaviour, so different asymptotic behaviour, which is discussed in Section \ref{Sannino}. This makes those potentials quantitatively different from the Starobinsky model in the context of eternal inflation and suggest, that eternal inflation cannot take place in those models. To be concrete, we have checked that eternal inflation for model described by \ref{Veff_RS} takes place at $\Phi \approx 1000\,M_{Pl}$, which is far beyond the applicability of the model. So now we turn to the inflationary models stemming from the asymptotic safety.
\FloatBarrier
\noindent
\FloatBarrier
\subsection{RG-improved Starobinsky inflation}
Renormalization Group improvement is a procedure of identifying and replacing the RG scale $k^2$ with a physical scale. It incorporates leading-order quantum effects in the dynamics of classical system. In the case of gravity, running of coupling constants in Einstein-Hilbert action results in additional contribution to the field equations from the gravitational energy-momentum tensor \cite{Platania:2020lqb}. In the de Sitter-type setting $k^2\sim R$ is the unique identification of the physical scale dictated by Bianchi identities \cite{Platania:2020lqb}. Such replacement in the scale-dependent Einstein-Hilbert action generates an effective $f(R)$ action, whose analytical expression is determined by running of the gravitational couplings. RG-improvement could solve classical black hole singularity problem \cite{Bonanno:1998ye,Bonanno:2006eu}, gives finite entanglement entropy \cite{Pagani:2018mke} and generates inflationary regime in quantum gravity \cite{eichhorn2019asymptotically}.\\ In this section, we study the asymptotically safe inflation based on RG-improved quadratic gravity Lagrangian, considered in \cite{Bonanno:2015fga,Platania:2020lqb}:
\begin{equation}
    \mathcal{L}_{k}=\frac{1}{16\pi g_{k}}\left(R-2\lambda_{k} k^2 \right)-\beta_{k} R^2,
\end{equation}
with the running dimensionless couplings $g_{k}$, $\lambda_{k}$, $\beta_{k}$ being the three relevant directions of the theory, with running given by \cite{Bonanno:2004sy}:
\beq
g_k= \frac{6\pi c_1 k^2}{6 \pi \mu^2 + 23 c_1(k^2-\mu^2)}, \quad \quad \beta_k = \beta_{\ast} + b_0 \left(\frac{k^2}{\mu^2}\right)^{-\theta_3/2},
\eeq
where $\mu$ is the infrared renormalization point such that $c_1=g_k(k=\mu)$ and $c_1$ and $b_0$ are integration constants. We introduce a parameter $\alpha$ as
\beq
\alpha=-2\mu^{\theta_3}b_0/M_P^2,
\eeq
that measures the departure from the non-gaussian fixed point (NGFP). One may find the behavior of the couplings near the  NGFP and substitute the appropriate expressions to the Lagrangian using the RG-improvement and the following identification of scale $k^2= \xi R$, where $\xi$ is an arbitrary parameter of order one.
\begin{figure}[t!]
\hfill
\subfigure[]{
\label{fig:PlataniaFit1}
\includegraphics[width=0.48\textwidth]{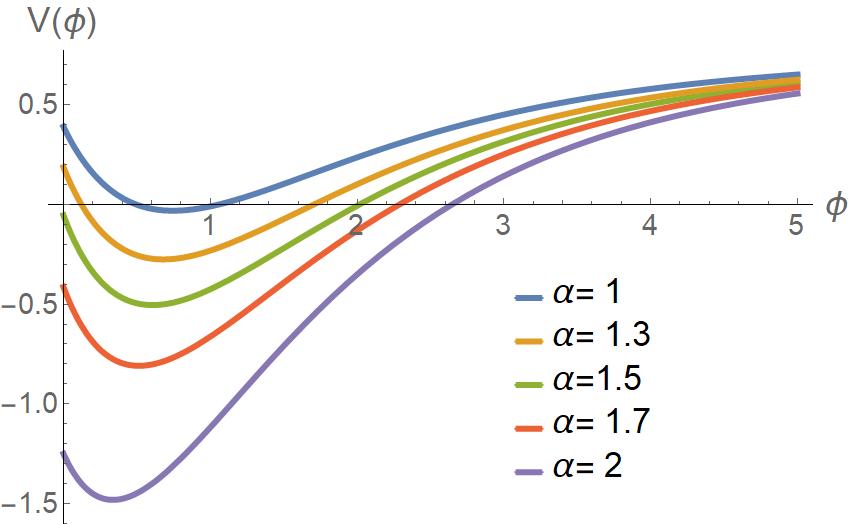}}
\hfill
\subfigure[]{
\label{fig:PlataniaFit2}
\includegraphics[width=0.48\textwidth]{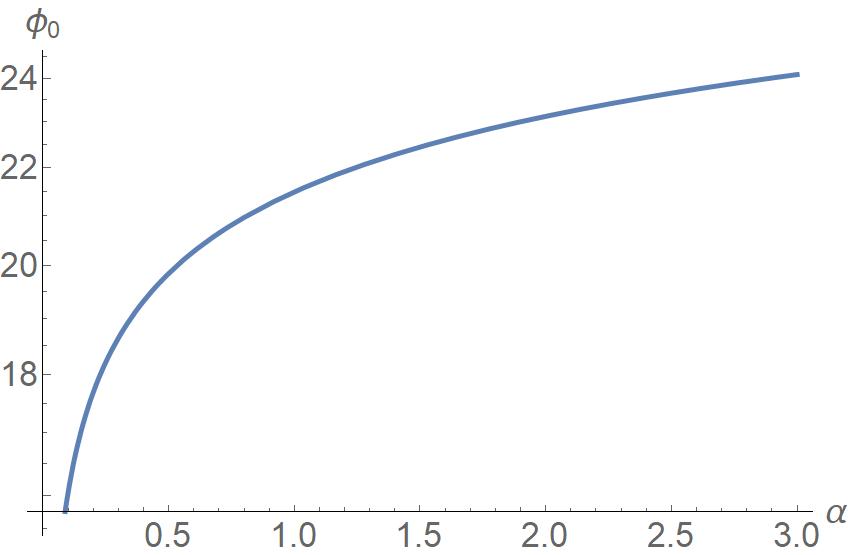}}
\hfill
\caption{Left: $V_{+}(\phi)$ plot for various $\alpha$ and fixed $\Lambda=1$ is shown. \newline
Right, logarithmic dependence of initial field value above which eternal inflation occurs on parameter $\alpha$ has been found. Blue points were evaluated via (\ref{eq:EternalCondition1}).}
\end{figure}
Following \cite{Bonanno:2015fga} we shall assume $\theta_3 =1$, then the transformation from the Jordan to the Einstein frame yields an effective potential \cite{Bonanno:2015fga,Platania:2020lqb}:
\begin{align}
\label{Vpm}
\begin{split}
    V_{\pm} =& \frac{m^2 e^{-2 \sqrt{\frac{2}{3}}\kappa\phi}}{256\kappa}\Bigg\{\vphantom{6 \alpha ^3  \sqrt{\alpha^2 + 16 e^{\sqrt{\frac{2}{3}}\kappa\phi}-16}} 192(e^{\sqrt{\frac{2}{3}}\kappa\phi}-1)^2 - 3\alpha ^4 + 128 \Lambda \\
    &- \sqrt{32} \alpha \left[ (\alpha ^2 + 8 e^{\sqrt{\frac{2}{3}}\kappa\phi} - 8) \pm \alpha \sqrt{\alpha ^2 + 16 e^{\sqrt{\frac{2}{3}}\kappa\phi} -16}\right]^{\frac{3}{2}}\\
    &- 3\alpha ^2 (\alpha ^2 + 16 e^{\sqrt{\frac{2}{3}}\kappa\phi} - 16) \mp 6 \alpha ^3  \sqrt{\alpha^2 + 16 e^{\sqrt{\frac{2}{3}}\kappa\phi}-16} \Bigg\}
\end{split},
\end{align}
where the only free parameters are cosmological constant $\Lambda$ and $\alpha$ after the CMB normalization we perform below. The $V_{+}$ branch predicts the reheating phase, figure \ref{fig:PlataniaFit1} shows its plot for various $\alpha$.\\
Similarly as in the case of Starobinsky inflation, we have denoted $V_0$ the constant part of the potential at infinity $V(\phi\xrightarrow{}\infty)=V_0=\frac{3m^2}{4 \kappa^2}$ and fixed it with CMB data by the relation (\ref{cmb}). For example, given $\alpha=2.8$, $\Lambda=1$ the plateau value is equal to $V_0=1.99\times10^{-10}$ $M^4_{Pl}$, hence one may fix the mass parameter $m=2\times10^{14}$ GeV.\\
Now we shall investigate the eternal inflation conditions given by (\ref{eq:EternalCondition1}, \ref{eq:EternalCondition2}). These conditions restricts the initial value of the field. We search for $\phi_0$ above which the eternal inflation occurs, as a function of the theory parameters. We have also found, that initial value above which eternal inflation occurs does not depend on the cosmological constant. It is due to the fact that $\Lambda$ only shifts the minimum of the potential and does not affect the large-field behavior of the system. Analytical conditions for EI have been checked for a set of $\alpha$ and depicted on figure \ref{fig:PlataniaFit2}.
The initial value of the field, depends logarithmically on $\alpha$.
%: $\phi_0(\alpha)=a\log(b\alpha)$, where the values $a=2.38$ $M_{Pl}$, $b=8917$. 
The reason for that behaviour is the following. In the large field expansion:
\begin{align}
V_{\pm}(\phi)=V_{plateau}-128 V_0  \alpha e^{-\frac{1}{2}\sqrt\frac{3}{2}\phi},
\end{align}
and by the substitution $\tilde{\phi} = e^{-\frac{1}{2}\sqrt\frac{3}{2}\phi}$ the potential reduces to the linear hilltop model, which justifies the usage of formulae (\ref{eq:EternalCondition1}, \ref{eq:EternalCondition2}) and the functional form of  $\phi_0(\alpha)$. The results were also confirmed by the numerical simulations. For example, given $\Lambda=1, \, \alpha=1.6$, the analytical considerations predict $\phi_{EI}=22.6 \, M_{Pl}$. The direct numerical simulation for this set of parameters yields $\Gamma=0.0001$ $M_{Pl}$, and $3 H=0.0003$ $M_{Pl}$, meaning that the eternal inflation begins slightly below the expected value $\phi_{EI}$. The plateau of (\ref{Vpm}) at large field values is a characteristic feature of effective inflationary potentials stemming from the asymptotically safe theories. It is dominated by eternal inflation and may suggest a deeper relation between the asymptotic safety of quantum gravity and multiverse. 
\subsection{Large N-dynamics and (eternal) inflation} %to jest sannino
\begin{figure}[t!]
\hfill
\subfigure[]{
\label{fig:RContour}
\includegraphics[width=0.48\textwidth]{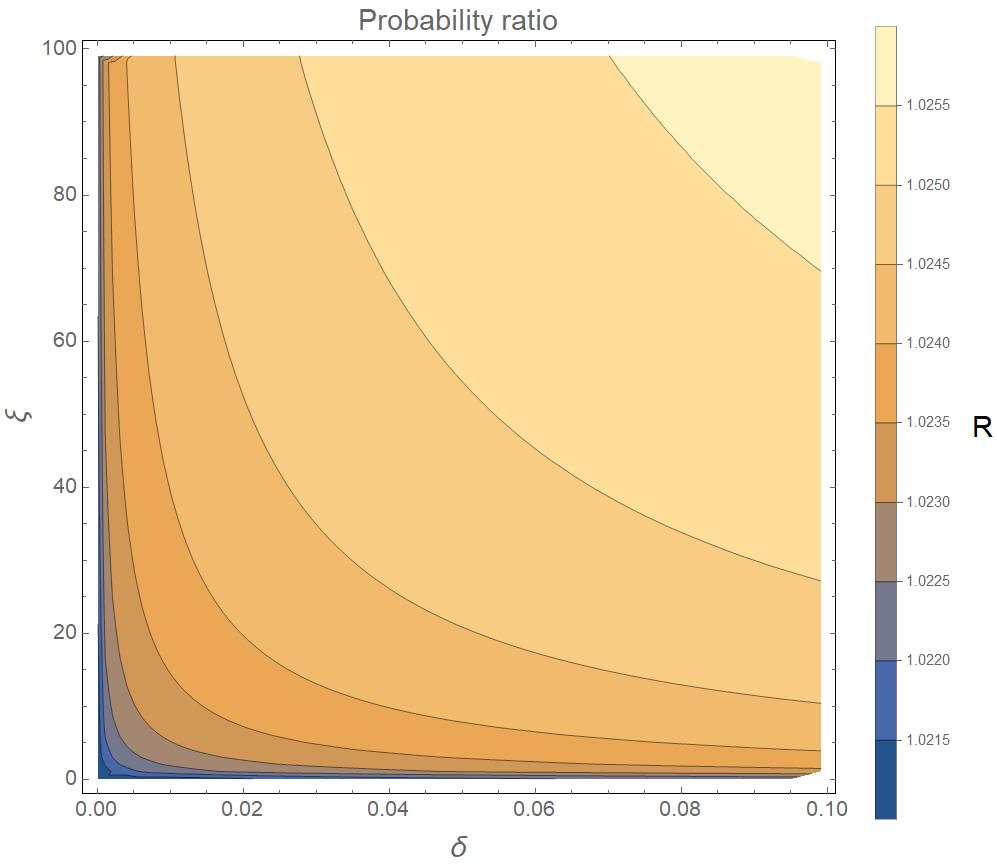}}
\hfill
\subfigure[]{
\label{fig:PhimaxContour}
\includegraphics[width=0.48\textwidth]{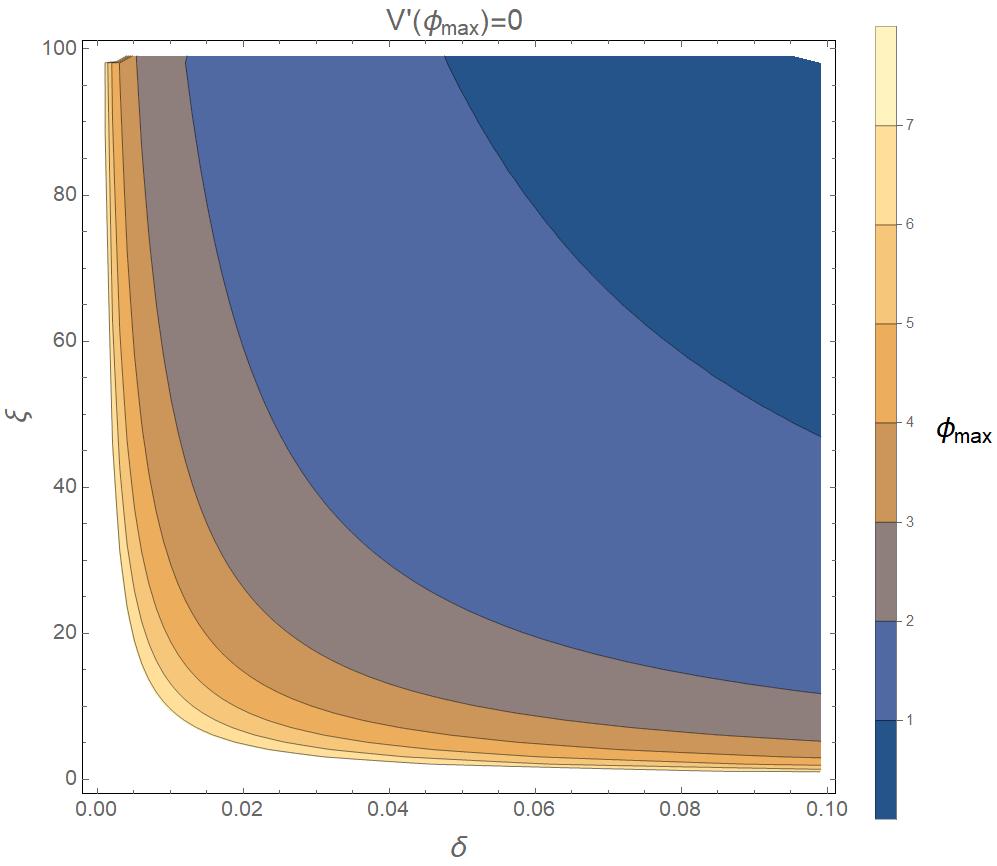}}
\hfill
\caption{Left: probability ratio $R=\frac{p_{+}}{p_{-}}$ of descending from the maximum towards the right minimum ($p_{+}$) and the left minimum ($p_{-}$), as a function of theory parameters $\xi$ and $\delta$. For the small values of the parameters it is more probable to fall from the maximum towards $\phi_{-}=0$ with non-eternal inflation, while for the large values of parameters, minimum at $\phi_{+}=\infty$ is favored, resulting in eternally inflating universe.\newline
Right: the value of the field at which, the potential is maximal. The above figures are qualitatively similar because for the small values of $\phi_{max}$, the effective potential is highly asymmetric ("step-like"). This brakes the symmetry between the right and left descend probability.}
\end{figure}

\begin{figure}[t!]
\hfill
\subfigure[]{
\label{fig:TunnelingProbabilities}
\includegraphics[width=0.48\textwidth]{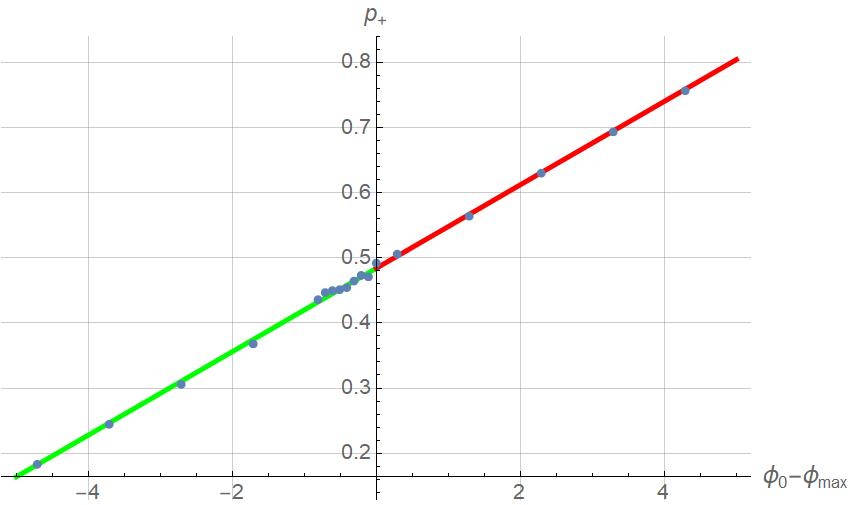}}
\hfill
\subfigure[]{
\label{fig:Rmax}
\includegraphics[width=0.48\textwidth]{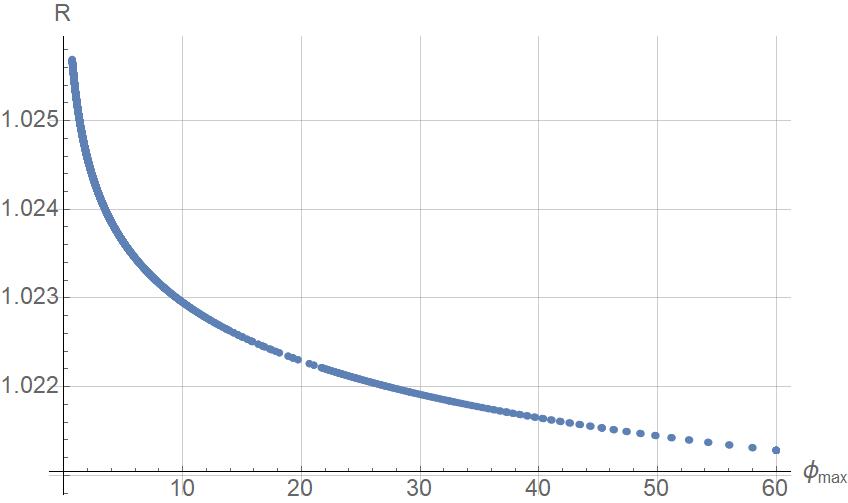}}
\hfill
\caption{Left: Linear probability distribution of tunneling (green side) and rolling (red side) towards the minimum at infinity as. The data points have been directly simulated.\newline
 Right: Probability ratio $R$, evaluated with (\ref{Rmaximum}), is a monotonically decreasing with the value of the maximum of the potential in the steepest descent approximation}
\end{figure}

\label{Sannino}
In this section we investigate model in which inflation is driven by an ultraviolet safe and interacting scalar sector stemming from a new class of non-supersymmetric gauge field theories. We consider a $\mathrm{SU}(N_C)$ gauge theory, with $N_F$ Dirac fermions and interacting with an $N_F$ $\times$ $N_F$ complex scalar matrix $H_{ij}$ that self interacts, described in \cite{Nielsen:2015una}. The Veneziano limit ($N_F \to +\infty$, $N_C\to +\infty$, $N_F/N_C=\mathrm{const}$) is taken such that the ratio $N_F/N_C$ becomes a continuous parameter \cite{Litim_2014}. 
The action in Jordan frame has the following form:
\begin{equation}
    S_J=\int d^4x \sqrt{-g} \left\{ - \frac{M^2+\xi\phi^2}{2}R + \frac{g^{\mu \nu}}{2}\partial_{\mu}\phi \partial_{\nu}\phi -   V_{\mathrm{iUVFP}} \right\},
\end{equation}
where the leading logarithmically resummed potential $V_{iUVFP}$ is given by:
\begin{equation}
    V_{\mathrm{iUVFP}}(\phi)=\frac{\lambda_* \phi^4}{4 N_f^2 \left(1+W(\phi)\right)}\left(\frac{W(\phi)}{W(\mu_0)}\right)^{\frac{18}{13 \delta}},
\end{equation}
where $ \lambda_* = \delta \frac{16 \pi^2}{19}(\sqrt{20+6\sqrt{23}}-\sqrt{23}-1)$ is positive quartic coupling at the fixed point, $\phi$ is the real scalar field along the diagonal of $H_{ij} = \phi \delta_{ij}/\sqrt{2N_f}$ and $\delta = N_F/N_C - 11/2$ is the positive control parameter, $W(\phi)$ is the Lambert function solving the transcendent equation 
\begin{equation}
    z = W \exp W, 
\end{equation}
with
\begin{equation}
    z(\mu) = \left(\frac{\mu_0}{\mu}\right)^{\frac{4}{3}\delta \alpha*}\left(\frac{\alpha*}{\alpha_0}- 1 \right) \exp \left[\frac{\alpha*}{\alpha_0} - 1 \right] .
\end{equation}    
The parameter $\alpha* = \frac{26}{57}\delta + O(\delta^2)$  is the gauge coupling at its UV fixed point value and $\alpha_0 = \alpha(\mu_0)$ is the same coupling at a reference scale $\mu_0$. \\ 
A conformal transformations allows to rewrite the action from Jordan to Einstein frame. Assuming a single field slow-roll inflation, we examine inflationary predictions of the potential and compute the slow-roll parameters:
\begin{equation}
    \epsilon = \frac{M_{Pl}^2}{2}\left(\frac{dU/d\chi}{U}\right)^2, \quad \quad
    \eta = M_{Pl}^2\frac{d^2U/d\chi^2}{U},
\end{equation}
where $U = V_{\mathrm{iUVFP}}/ \Omega^4$, with $\Omega^2 = (M^2 + \xi \phi^2)/M_{Pl}^2$ being the conformal transformation of the metric, and $\chi$ is the canonically normalized field in the Einstein frame. We assume, that $M = M_{Pl}$. Inflation ends when the slow-roll conditions are violated, that is when $\epsilon(\phi_{end})$ or |$\eta(\phi_{end})$| = 1. We analyze the non-minimal case, where the coupling $\xi$ is non-vanishing. The potential $U$ is given by:
\begin{equation}
\label{NonMinimalPotential}
    U=\frac{V_{\mathrm{iUVFP}}}{\Omega^4} \approx \frac{\lambda_* \phi^4}{4 N_F^2 \left( 1+ \frac{\xi \phi^2}{M_{Pl}^2} \right)^2} \left(\frac{\phi}{\mu_0} \right)^{-\frac{16}{19}\delta}
\mathrm{.}\end{equation}
\begin{figure}[t!]
\hfill
\subfigure[]{
\label{fig:Sanninoplot}
\includegraphics[width=0.48\textwidth]{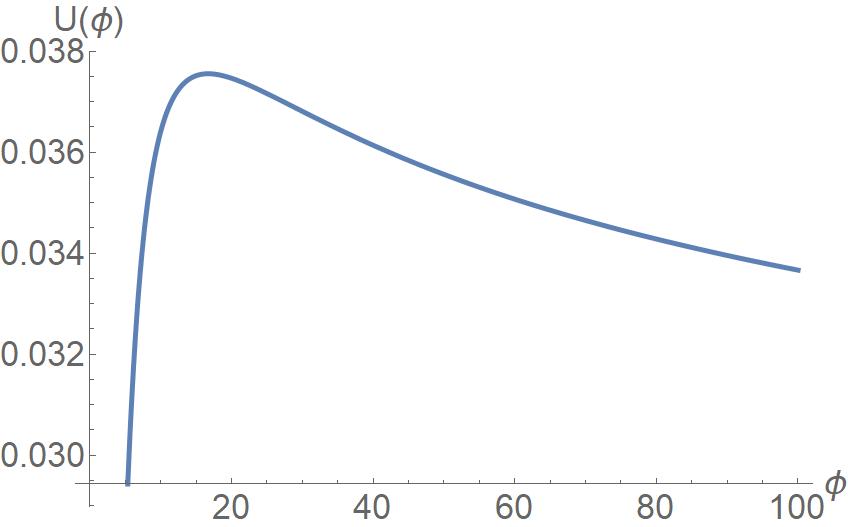}}
\hfill
\subfigure[]{
\label{fig:SanninoCondition1}
\includegraphics[width=0.48\textwidth]{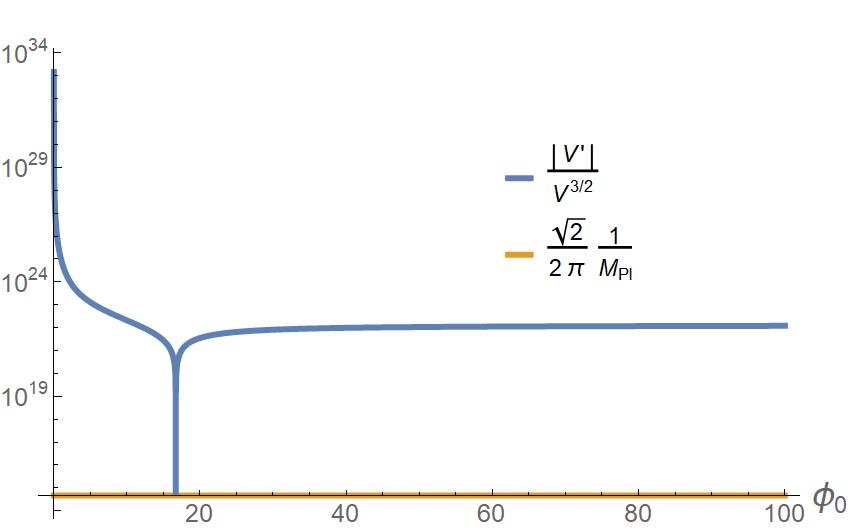}}
\hfill
\caption{Left: the non-minimally coupled potential as a function of $\phi$ for $\delta = 0.1$, $\xi = 1/6$ and $\mu_{0}= 10^{-3}M_{Pl}$. There is a maximum at $\phi_{max}=16.7\, M_{Pl}$.\newline
Right: For the same set of parameters, we plot the first eternal inflation condition as a function of $\phi$ (blue curve) and the eternal inflation bound (yellow curve). Inflation becomes eternal if the blue curve is below the yellow one. At $\phi_{max}$ the first derivative of the potential vanishes and (\ref{eq:EternalCondition1}) predicts a narrow window for the eternal inflation.}
\end{figure}
In the large field limit $\phi$ $\gg$ $M_{Pl}/\sqrt{\xi}$ the $\phi^4$ term in the numerator cancels against the term in the denominator. In this limit, the quantum corrections dictate the behaviour of the potential, which is found to decrease as:
\begin{equation}\label{eqn:maxima}
    \frac{\lambda_* M_{Pl}^4}{4 N_F^2 \xi^2}\left(\frac{\phi}{\mu_0}\right)^{-\frac{16}{19}\delta}
\mathrm{.}\end{equation}
The non-minimal coupled potential has one local maximum and two minima. The region to the left of the maximum is the region, where the inflation can be brought to an end and the reheating takes place \cite{Svendsen:2016kvn}. To the right of the maximum, the inflation becomes classically eternal. For large values of $\phi$ the potential flattens out and the slow-roll conditions are not violated. Numerical solutions to the FP equation shows that the is no possibility of eternal inflation in that region, since it is an unstable maximum and hence any quantum fluctuation will drop it from that position. Furthermore due to steepness of the potential around this maxima there is no possibility for the field to remain in that region.
Let us now investigate the analytical eternal inflation conditions. Similarly as in the Starobinsky model, the second condition (\ref{eq:EternalCondition2}) is always satisfied. The first condition (\ref{eq:EternalCondition1}) is illustrated on the figure \ref{fig:SanninoCondition1}. There is a peak for $\phi$ = $\phi_{max}$ = 16.7 $M_{Pl}$, due to the vanishing derivative and if we "zoom in", the analytical condition allows for eternal inflation in the close neighbourhood of $\phi_{max}$. We have verified numerically this is not a sustainable attractor of eternal inflation. A field that starts evolution at $\phi_{max}$ will leave this region, as it cannot climb further up-hill. Nevertheless, eternal inflation may still occur due to the quantum tunnelling through the potential barrier.
\paragraph{Tunneling through the potential barrier}
\label{tunneling section}
As described in section \ref{SecTunneling}, if the potential has multiple vacua, quantum tunneling through the potential barrier is expected. The non-minimal coupling potential (\ref{NonMinimalPotential}) belongs in this class. The question is, whether tunnelling from the non-eternal inflation region of $\phi<\phi_{max}$ to the region of classical eternal inflation $\phi>\phi_{max}$ is possible.\\
We start with investigating the fate of the field initially placed at the peak $\phi_0=\phi_{max}$ of the potential depicted on figure \ref{fig:Sanninoplot}. By the virtue of the steepest descent approximation at the maximum equation (\ref{Rmaximum}) may be employed. The resulting ratio of probabilities of the right-side descent to the left-side descent $R(\alpha,\xi)=\frac{p_{+}}{p_{-}}$, as a function of parameter $\alpha$ and the non-minimal coupling constant $\xi$ was calculated directly from the formula (\ref{Rmaximum}) without the need of numerical simulation of the Langevin equation. It is presented on figure \ref{fig:RContour}. Due to the complexity of the potential (\ref{NonMinimalPotential}) its maximum was found numerically and then employed in (\ref{Rmaximum}). Figure \ref{fig:PhimaxContour} shows how the maximum changes with parameters. As expected, the ratio $R(\alpha,\xi)$ is close to 1 and favors the right-side (left-side) of the potential for large (small) values of the parameters. The biggest ratio emerges at large values of the parameters $\xi$ and $\delta$, since the potential is "step-like" and highly asymmetric. It is monotonically decreasing with the values of $\phi_{max}$, for which the potential has a maximum, see figure \ref{fig:Rmax}.
\FloatBarrier
\FloatBarrier
In order to verify the accuracy of the relation (\ref{Rmaximum}), we have performed numerical simulation of the discretized Langevin equation (\ref{DiscretLangevin}) with initial condition $\phi_0=\phi_{max}$. For example it was found, that for the set of the parameters $N_F=10$, $\mu=10^{-3} M_{Pl}$, $\xi=\frac{1}{6}$, $\delta=0.1$ the steepest descent approximation yields $R=0.92$, and the numerical analysis results in $R=0.97$, which proves good accuracy of the analytical formula.\\ 
One may wonder, how does the probability $p_{\pm}(\phi_0)$ depends on the departure from the maximum $\phi_0 \neq \phi_{max}$. The analytical answer is given by (\ref{AnalP}). As we have checked numerically inflation becomes eternal, when tunneling probability is non-zero, as depicted on Figure \ref{fig:TunnelingProbabilities}.\\  
To bypass the numerical calculation of the integral (\ref{AnalP}) we employ the direct numerical simulation of the Langevin equation. This time however, we do not seek for the time evolution of the inflaton. Rather than creating histograms of the count of inflationary events at a given timestep, we simply track the probabilities $p_{+}$ and $p_{-}$. We say that the particle tunnelled through the potential barrier contributing to $p_{+}$ if the evolution starts at $\phi_0<\phi_{max}$ and proceeds to arbitrarily large field values after a long time. For each point at figure \ref{fig:TunnelingProbabilities} the probability has been calculated on the sample of 10000 simulations. As expected, choosing values of $\phi_0$ smaller than $\phi_{max}$ lowers the probability of the tunneling to the right side of the barrier. Moreover, the probability of tunneling decreases linearly with the distance to the maximum. The result of the simulation for the set of parameters 
\beq
\label{eq:Sanninopar}
N_F=10,\quad \mu=10^{-3} M_{Pl},\quad \xi=\frac{1}{6},\quad \delta=0.1
\eeq
is shown on figure \ref{fig:TunnelingProbabilities}. The green line corresponds to the green ball on figure \ref{fig:Vacua1} and shows the probability of tunneling through the barrier (as in figure \ref{fig:Vacua2}) as a function of proximity to the maximum $\phi_0 \neq \phi_{max}$. The red line corresponds to the red ball on figure \ref{fig:Vacua1} and shows the probability of rolling towards the minimum at infinity. \\
The rolling is also a stochastic process, as the tunneling in the opposite direction is possible. The probability distribution of tunneling in either direction is not a symmetric process. Notice, that the initial condition, for which $p_{+}=\frac{1}{2}$ is shifted to the right of $\phi_{max}$. This means, that starting from the maximum, it is slightly more probable to land in $\phi_{-}$. There is a point, below which the green ball cannot tunnel $p_{+}=0$ (for the set of parameters given by (\ref{eq:Sanninopar}), at $9.2 M_{Pl}$), and the other limiting case (at $24.8$ $M_P$), when the red ball cannot tunnel $p_{-}=0$. Hence, for every initial value of the field above $9.2$ Planck Masses, there is a non-zero probability of eternal inflation. On the other hand, for $\phi_0=9.2 \,M_{Pl}$ and parameters given by (\ref{eq:Sanninopar}), the inflation classically produces roughly 54 e-folds, depending on the reheating time \cite{Svendsen:2016kvn}, and is in the agreement with CMB data. This shows that the model is on a verge of being eternally inflating, which may point out to the interesting phenomenology.\\
To sum up the critical point of our analysis is that the analytical conditions (\ref{eq:EternalCondition1}, \ref{eq:EternalCondition2}) did not allow for eternal inflation, even though the tunneling process may evolve any initial point above 9.2 Planck Masses to $\phi_{+}=\infty$, and not violate the slow-roll conditions. This shows that conditions (\ref{eq:EternalCondition1},\ref{eq:EternalCondition2}) are not well suited for the multiple minima models and cannot contain the full information of the global influence of quantum fluctuations in the early universe.
\FloatBarrier
\section{Conclusions}
\label{Sec:Discussion}
Eternal inflation remains a conceptual issue of the inflationary paradigm. The creation of scattered, causally disconnected regions of spacetime - the multiverse is not confirmed observationally and raises question about the inflationary predictions \cite{Ijjas:2014nta}. Hence, one may impose the no eternal inflation principle \cite{Rudelius:2019cfh} to restrict free parameters and the initial conditions.\\
We have investigated popular inflationary models, and have found that in principle, eternal inflation is present at every asymptotically flat effective potential for large field values, assuming the ergodicity of the system. The finite inflationary time of our pocket universe serves as a consistency condition of the multiverse predictions. In section \ref{AlfaAttractorSection}, we verified that $\alpha$-attractor T-models are consistent from this point of view.\\
If the initial value of the scalar field driving inflation is above the Planck scale, a UV-completeness of a given model is necessary. Starobinsky inflation stemming from $R^2$ gravity gives around 60 e-folds for $\phi_0$=5.5 $M_{pl}$. We have considered
the effective quantum corrections to the Starobinsky inflation based on the qualitative behavior of the running coupling constants. Next, the RG-improvement of $R+R^2$ Lagrangian was studied in this case and we have found that field values required for eternal inflation are typically higher than the ones for the Starobinsky case. The flatness of the potential and possibility of eternal inflation seems to be a signature mark of asymptotically safe UV completions, in contradistinction to the effective theory corrections. We have checked that \cite{Liu:2018hno} $\Phi \sim 1000$ $M_{Pl}$ in order to get eternal inflation, which is far beyond the applicability of the model. \\
Furthermore we have found, that for potentials with multiple vacua, tunneling through potential barriers provides a new mechanism for eternal inflation. So in order to understand the inflationary dynamics one cannot simply cut the potential at the maximum. The $\mathrm{SU}(N)$ Gauge theory with Dirac fermions provides an example for such behavior. The probability of tunneling to the side dominated by eternal inflation becomes negligible few Planck Masses away from the peak of the potential. Yet the fixed point values of the couplings and possibly the shape of the potential can be obscured by the quantum gravity effects and this shall be investigated elsewhere.\\
Our analysis reveals that there is no obstruction for the multiverse scenario in the asymptotically safe models. Yet its occurrence depends on the initial conditions for the inflationary phase and the matching to the observational data, tying these three profound issues together. On the other hand, in AS models these questions can have intriguing answers by the finite action principle \cite{Lehners:2019ibe}.

\acknowledgments We thank J. Reszke and J. Łukasik for participating in the early stages of this project. We thank G. Dvali, A. Eichhorn, M. Pauli, A. Platania, T. Rudelius, S. Vagnozzi and Z.W. Wang for fruitful discussions and extensive comments on the manuscript. Work of J.H.K. was supported by the Polish National Science Center (NCN) grant 2018/29/N/ST2/01743. J.H.K. would like to acknowledge the CP3-Origins hospitality during this work. The computational part of this research has been partially supported by the PL-Grid Infrastructure.
%\bibliographystyle{unsrt}
%\bibliography{References.bib}
\bibliography{output (3).bbl}
\end{document}